\documentclass[aps,prl,twocolumn,groupedaddress]{revtex4}
\begin{document}

\title{Scaling Limit of Vicious Walkers, Schur Function, \\
and Gaussian Random Matrix Ensemble}

\author{Makoto Katori}
\email[]{katori@phys.chuo-u.ac.jp}
\altaffiliation{On leave from Department of Physics,
Faculty of Science and Engineering,
Chuo University, Kasuga, Bunkyo-ku, Tokyo 112-8551, Japan}
\affiliation{
University of Oxford, 
Department of Physics--Theoretical Physics, 
1 Keble Road, Oxford OX1 3NP, United Kingdom}
\author{Hideki Tanemura}
\email[]{tanemura@math.s.chiba-u.ac.jp }
\affiliation{
Department of Mathematics and Informatics,
Faculty of Science, Chiba University, 1-33 Yayoi-cho, Inage-ku,
Chiba 263-8522, Japan}

\date{\today}

\begin{abstract}
We consider the diffusion scaling limit of the vicious walkers
and derive the time-dependent spatial-distribution
function of walkers. The dependence on initial configurations
of walkers is generally described by using the symmetric
polynomials called the Schur functions. In the special
case in the scaling limit that all walkers are
started from the origin, the probability density
is simplified and it shows that the positions of walkers 
on the real axis at time one is identically
distributed with the eigenvalues of random matrices
in the Gaussian orthogonal ensemble.
Since the diffusion scaling limit makes the vicious walkers
converge to the nonintersecting Brownian motions
in distribution, the present study will provide a new method
to analyze intersection problems of 
Brownian motions in one-dimension.
\\
\nonumber PACS numbers: 05.40.-a, 05.50.+q, 02.50.Ey
\end{abstract}


\maketitle

The problem of vicious walkers was introduced by Fisher
and its application to various wetting and melting phenomena
were described in his Boltzmann medal lecture \cite{Fis84}.
Recently, by using the standard one-to-one correspondence
between walks and Young tableaux,
Guttmann {\it et al.} \cite{GOV98} and 
Krattenthaler {\it et al.} \cite{KGV00} showed that the
exact solutions for some enumeration problems of vicious
walks \cite{AME91,EG95} are derived from
the theory of symmetric functions \cite{Mac95}
or the representation theory of classical groups \cite{FH91}.
Important analogies between the ensembles of Young tableaux
and those of Gaussian random matrices were reported by 
Johansson \cite{Joh00}, and Baik \cite{Bai00} and 
Nagao and Forrester \cite{NF01} studied the vicious walker
problem using the random matrix theory of the
Gaussian orthogonal ensemble (GOE) \cite{Meh91,Dei00}.

The purpose of this Letter is to demonstrate more explicit
relations among the vicious walker problem, 
the Schur function \cite{Schur}, and
the GOE, by performing the diffusion scaling limit 
of the vicious walkers.
We derive the time-dependent spatial-distribution 
function of walkers in this scaling limit.
The dependence on the initial configurations is generally
described by using the Schur functions.
We show that the case, in which all walkers
are started from the origin, can be treated, and in this
special case the probability density of positions of walkers
at time $t=1$ is identified with the probability density
of eigenvalues in the GOE.
Since the distribution of random walkers converges to that
of Brownian motions in the diffusion scaling limit,
the present analysis will solve some intersection problems
of one-dimensional Brownian motions.
More applications to the probability theory will be
reported elsewhere \cite{KT}

Vicious walks are defined as a subset of the simple random
walks as follows. Let $\{R_{k}^{s_{i}}\}_{k \geq 0},
i \in I_{n} \equiv \{1,2, \cdots, n\}$, 
be the $n$ independent symmetric simple random walks on 
${\bf Z}=\{ \cdots, -2, -1, 0, 1 , 2 \cdots \}$
started from $n$ distinct positions,
$2s_{1} < 2s_{2} < \cdots < 2s_{n}$,
$s_{i} \in {\bf Z}$.
Fix the time interval $m$ as a positive even number.
The total number of walks is $2^{mn}$,
all of which are assumed to be realized with equal
probability $2^{-mn}$.
Now we consider a subset of walks such that any of walkers
does not meet other walkers up to time $m$, that is,
the condition
\begin{equation}
R_{k}^{s_{1}} < R_{k}^{s_{2}} < \cdots
< R_{k}^{s_{n}} \quad 1 \leq \forall k \leq m
\label{eqn:vicious}
\end{equation}
is imposed.
Such a subset of walks is called the vicious walks
(up to time $m$) \cite{Fis84}.
Let $N_{n}(m; \{e_{i}\}| \{s_{i}\})$ be the total number
of the vicious walks, in which the $n$ walkers 
arrive at the positions
$2e_{1} < 2e_{2} < \cdots < 2e_{n}$ at time $m$.
Then the probability that such vicious walks with
fixed end-points are realized 
in all possible random walks started from
the given initial configuration is 
$N_{n}(m; \{e_{i}\} |\{s_{i}\})/2^{mn}$, 
which is denoted by $V_{n}(\{R_{k}^{s_{i}} \}_{k=0}^{m};
R_{m}^{s_{i}}=2 e_{i})$ in this Letter.
We also define
$$
V_{n}(\{R_{k}^{s_{i}} \}_{k=0}^{m}) 
=\sum_{e_{1} < e_{2} < \cdots < e_{n}}
V_{n}(\{R_{k}^{s_{i}} \}_{k=0}^{m};
R_{m}^{s_{i}}=2 e_{i}).
$$

Recently Krattenthaler {\it et al.} \cite{KGV00} 
evaluated the asymptotes for large $m$ 
of $V_{n}(\{R_{k}^{s_{i}}\}_{k=0}^{m})$
for the two special initial-configurations,
(i) $s_{i}=i-1$ and (ii) $s_{i}=2(i-1)$, as 
$$
V_{n}(\{R_{k}^{s_{i}}\}_{k=0}^{m})
= a_{n} b_{n}(\{s_{i}\}) m^{-n(n-1)/4}
\left( 1+ {\cal O}(1/m) \right), 
$$
where
\begin{eqnarray}
a_{n} &=&
\left\{
   \begin{array}{ll}
      (2^{n}/\pi)^{n/4} \
      \displaystyle{\prod_{i=1}^{n/2} (2i-2) ! }
      & \mbox{if} \ n=\mbox{even} \\
      (2^{n+1}/\pi)^{(n-1)/4}
      \displaystyle{\prod_{i=1}^{(n-1)/2} (2i-1) ! }
      & \mbox{if} \ n=\mbox{odd}, \\
   \end{array}\right.
\label{eqn:Krat2}
\end{eqnarray}
with
$
b_{n}(\{i-1\})=1, b_{n}(\{2(i-1)\})=2^{n(n-1)/2}.
$

We found that their result can be immediately 
generalized as
\begin{equation}
 b_{n}(\{s(i-1)\})=s^{n(n-1)/2} \
\mbox{for any} \ s=1,2, \cdots.
\label{eqn:Krat3}
\end{equation}
This observation suggests that we can take the 
scaling limit
such that $L \to \infty$ with the time interval
$m=L t$ and the initial spacing of walkers 
$s = \sqrt{L}/2$,
where $t$ is finite.

In this Letter we consider the 
{\it diffusion scaling limit};
setting $m=Lt, s_{i}=\sqrt{L} x_{i}/2,
e_{i}=\sqrt{L} y_{i}/2, i \in I_{n}$, and taking 
the limit $L \to \infty$.
The key lemma, which will be proved shortly, is the
following.
For given $t >0$ and 
$0 \leq x_{1} < x_{2} < \cdots < x_{n} \
(x_{i} \in {\bf Z}), 
y_{1} < y_{2} < \cdots < y_{n}$,
let $\xi(x)=(\xi_{1}(x), \cdots, \xi_{n}(x))$ 
be a partition specified by the starting positions
$\{x_{i}\}$ as
\begin{equation}
 \xi_{i}(x)=x_{n-i+1}-(n-i), \
 i \in I_{n},
\label{eqn:xi}
\end{equation}
then
\begin{eqnarray}
&& \lim_{L \to \infty} \left(\frac{\sqrt{L}}{2}\right)^{n} 
V_{n} \left(
\left\{R_{k}^{\sqrt{L} x_{i} /2} \right\}_{k=0}^{Lt};
R_{Lt}^{\sqrt{L} x_{i}/2}=
\frac{\sqrt{L} y_{i}}{2} \right)
\nonumber\\
&&  = (2 \pi t)^{-n/2} 
s_{\xi(x)}\left(e^{y_{1}/t}, e^{y_{2}/t}, \cdots,
e^{y_{n}/t}\right) \nonumber\\
&& \ \times \exp\left(- \frac{1}{2t} \sum_{i=1}^{n}
(x_{i}^2+y_{i}^2) \right)
\prod_{1 \leq i < j \leq n}
(e^{y_{j}/t}-e^{y_{i}/t}),
\label{eqn:Lemma}
\end{eqnarray}
where $s_{\lambda}(z_{1}, \cdots, z_{n})$ is
the Schur function \cite{Schur}.
We consider the rescaled one-dimensional lattice
${\bf Z}/(\sqrt{L}/2)$, where the unit length is 
$2/\sqrt{L}$,
and let $\tilde{R}_{k}^{x}$ denote the symmetric simple
random walk starting from $x$ on ${\bf Z}/(\sqrt{L}/2)$.
Then (\ref{eqn:Lemma}) implies that
\begin{eqnarray}
&& \lim_{L \to \infty}
V_{n}\left(\left\{ \tilde{R}_{k}^{x_{i}} 
\right\}_{k=0}^{Lt};
\tilde{R}_{Lt}^{x_{i}} \in [y_{i}, y_{i}+dy_{i}]
\right) \nonumber\\
&& \qquad = f_{n}(t; \{y_{i}\}|\{x_{i}\}) d^{n} y,
\nonumber
\end{eqnarray}
where $f_{n}(t; \{y_{i}\}|\{x_{i}\})$ is defined by
the RHS of (\ref{eqn:Lemma}).

In order to normalize $f_{n}$, we consider the integral
\begin{eqnarray}
&&{\cal N}_{n}(t; \{x_{i}\}) =
\int_{y_{1} < \cdots < y_{n}} d^{n}y \
f_{n}(t; \{y_{i}\}|\{x_{i}\}) \nonumber\\
&& = \frac{e^{-\sum x_{i}^2/2t}}{(2 \pi t)^{n/2} n !} \
\int d^{n}y \
s_{\xi(x)}(e^{y_{1}/t}, \cdots, e^{y_{n}/t}) \nonumber\\
&&  \quad \qquad \times e^{-\sum y_{i}^2/2t} 
\prod_{1 \leq i < j \leq n}|e^{y_{j}/t}-e^{y_{i}/t}|,
\label{eqn:norm}
\end{eqnarray}
where we have used the fact that with the absolute values
the integrand is invariant under permutation of $y_{i}$,
since the Schur function is a symmetric function.
The spatial-distribution function in the scaling limit 
of the $n$ vicious walkers at finite time $t$
is then given by 
$d\mu_{n}=f_{n} d^{n}y /{\cal N}_{n}$,
or more explicitly
$$
d \mu_{n}(t; \{y_{i}\} | \{x_{i}\}) =
g_{n}(t; \{y_{i}\} | \{x_{i}\}) d^{n}y,
$$
with the probability density
\begin{eqnarray}
&& g_{n}(t; \{y_{i}\} | \{x_{i}\}) 
= \frac{{\bf 1}\{y_{1} < y_{2} < \cdots < y_{n}\}}
{Z_{n}(t; \{x_{i}\}) } \nonumber\\
&& \quad \times s_{\xi(x)}(e^{y_{1}/t}, \cdots, e^{y_{n}/t})
\nonumber\\
&& \quad \times
\exp \left(-\frac{1}{2t} \sum_{i=1}^{n}y_{i}^2 \right)
\prod_{1 \leq i < j \leq n} \
(e^{y_{j}/t}-e^{y_{i}/t})  
\label{eqn:dmu}
\end{eqnarray}
for $0 \leq x_{1} < \cdots < x_{n},
x_{i} \in {\bf Z}$, 
where ${\bf 1}\{\omega\}$ is 1 if $\omega$ is
satisfied and is zero otherwise, and
$
Z_{n}(t; \{x_{i}\}) 
=(2 \pi t)^{n/2} e^{\sum x_{i}^2/2t}
{\cal N}_{n}(t; \{x_{i}\}).
$
It should be noted that, if the initial configuration is
$x_{i}=i-1, i \in I_{n}$, then 
$\xi_{i}(x) \equiv 0$ and the Schur function 
in (\ref{eqn:norm}) and (\ref{eqn:dmu}) is
$s_{\xi(x)}=1$.

Now we give a proof of (\ref{eqn:Lemma}).
Define a subset of the square lattice ${\bf Z}^{2}$,
$$
{\cal L}_{m}=\{(x,y) \in {\bf Z}^{2}:
x+y=\mbox{even}, \ 0 \leq y \leq m \},
$$
and ${\cal E}_{m}$ be the set of all edges which connect the
nearest-neighbor pairs of vertices in ${\cal L}_{m}$.
Then each walk of the $i$-th walker, $i \in I_{n}$,
can be represented as a sequence of successive edges
connecting vertices $S_{i}=(2s_{i}, 0)$ and $E_{i}=(2e_{i}, m)$ 
on $({\cal L}_{m}, {\cal E}_{m})$, which we call the {\it lattice path}
running from $S_{i}$ to $E_{i}$.
If such lattice paths share a common vertex, they are said to 
intersect. Under the vicious walk condition
(\ref{eqn:vicious}), what we consider is a set of all 
$n$-tuples of {\it nonintersecting paths} \cite{Stem90}.
Let $\pi(S \to E)$ be
the set of all lattice paths from $S$ to $E$,
and
$\pi_{0}(\{S_{i}\}_{i=1}^{n} \to \{E_{i}\}_{i=1}^{n})$
be the set of all $n$-tuples $(\pi_{1}, \cdots, \pi_{n})$ of 
nonintersecting lattice
paths, in which $\pi_{i}$ runs from $S_{i}$ to $E_{i}$,
$i \in I_ {n}$.
If we write the number of elements in a set $A$ as
$|A|$, then
$N_{n}(m;  \{e_{i}\} | \{s_{i}\})=
|\pi_{0}(\{S_{i}\}_{i=1}^{n} \to \{E_{i}\}_{i=1}^{n})|
$ and the Lindstr\"om-Gessel-Viennot theorem
gives \cite{GV85,Stem90},
$$
N_{n}(m; \{e_{i}\} | \{s_{i}\})
= \det_{1 \leq i, j \leq n}
( |\pi(S_{j} \to E_{i})|).
$$
Since
$
|\pi(S_{j} \to E_{i})|=
{m \choose m/2+s_{j}-e_{i} },
$
we have the binomial determinant 
\begin{eqnarray}
&&V_{n}(\{R_{k}^{s_{i}}\}_{k=0}^{m};
R_{m}^{s_{i}}=2e_{i}) \nonumber\\
&& \quad= 2^{-mn} \det_{1 \leq i, j \leq n} 
\left( {m \choose m/2+s_{j}-e_{i}} \right).
\nonumber
\end{eqnarray}
Application of Stirling's formula yields
\begin{eqnarray}
&& \lim_{L \to \infty} 2^{-Ltn} (\sqrt{L}/2)^{n} 
\det_{1 \leq i, j \leq n}
\left( {Lt \choose Lt/2+\sqrt{L}(x_{j}-y_{i})/2} \right)
\nonumber\\
&& =
\det_{1 \leq i, j \leq n}
\left( \lim_{L \to \infty}  
2^{-Lt} (\sqrt{L}/2)  {Lt \choose 
Lt/2+\sqrt{L}(x_{j}-y_{i})/2} \right)
\nonumber\\
&& = 
\det_{1 \leq i, j \leq n}
\left( (2 \pi t)^{-1/2} \ e^{-(x_{j}-y_{i})^2/2t} \right) 
\nonumber\\
&& = 
(2 \pi t)^{-n/2} e^{-\sum (x_{i}^2+y_{i}^2)/2t}
\det_{1 \leq i, j \leq n}
\left( e^{x_{j} y_{i}/t} \right). \nonumber
\end{eqnarray}
We rewrite the determinant as
\begin{eqnarray}
&& \det_{1 \leq i, j \leq n}
\left( e^{x_{j} y_{i}/t} \right) \nonumber\\
&& = \frac{
\det_{1 \leq i, j \leq n} \left(
(e^{y_{i}/t})^{x_{n-j+1}} \right)}
{\det_{1 \leq i, j \leq n} \left(
(e^{y_{i}/t})^{n-j} \right)}
\times \Delta_{n}\left(
\{ e^{y_{i}/t} \} \right), \nonumber
\end{eqnarray}
where $\Delta_{n}(\{z_{i}\})$ is
the Vandermonde determinant
$$
\Delta_{n}(\{z_{i}\}) \equiv
\det_{1 \leq i, j \leq n} (z_{i}^{j-1})
= \prod_{1 \leq i < j \leq n}(z_{j}-z_{i}).
$$
Combining with (\ref{eqn:xi}) and the definition of
Schur function \cite{Schur}
completes the proof of (\ref{eqn:Lemma}).

Since 
\begin{eqnarray}
&&s_{\xi(\ell x)}(e^{y_{1}/\ell t}, \cdots,
e^{y_{n}/\ell t}) 
\Delta(\{e^{y_{i}/\ell t} \}) \nonumber\\
&&=s_{\xi}(x)(e^{y_{1}/ t}, \cdots,
e^{y_{n}/ t}) 
\Delta(\{e^{y_{i}} \})
=\det_{1 \leq i, j \leq n}
\left( e^{x_{j}y_{i}/t} \right)
\nonumber
\end{eqnarray}
for any integer $\ell$, where
$\ell x=(\ell x_{1}, \cdots, \ell x_{n})$,
we can prove the following {\it scaling property}
for the scaling-limit probability density.
For any integer $\ell$
\begin{equation}
g_{n}(t; \{y_{i}\}|\{x_{i}\})
=\ell^{n} g_{n}(\ell^2 t;
\{\ell y_{i}\}|\{\ell x_{i}\}).
\label{eqn:scaling}
\end{equation}
Using this property, we can generalize the expression
(\ref{eqn:dmu}) for any rational numbers,
$x_{1} < \cdots < x_{n}$, and then using the connectedness
of real numbers, for any real numbers $\{x_{i}\}$.

Next we study the $t \to \infty$ asymptotes
of the above results.
Since \cite{Schur}
\begin{eqnarray}
&& \lim_{t \to \infty} s_{\xi(x)}(e^{y_{1}/t}, 
\cdots, e^{y_{n}/t})
= s_{\xi(x)}(1,1, \cdots, 1) \nonumber\\
&& = \prod_{1 \leq i < j \leq n}
\frac{x_{j}-x_{i}}{j-i}
\equiv b_{n}(\{x_{i}\}) \nonumber
\end{eqnarray}
(remark that this definition of $b_{n}(\{x\})$ is
consistent with (\ref{eqn:Krat3})), and
$$ 
\lim_{t \to \infty} t^{n(n-1)/2}
\prod_{1 \leq i < j \leq n}
(e^{y_{j}/t}-e^{y_{i}/t})
= \prod_{1 \leq i < j \leq n}
(y_{j}-y_{i}),
$$
the normalization factor is asymptotically
\begin{eqnarray}
&& {\cal N}_{n}(t; \{x_{i}\}) =
t^{-n^2/2} \left(1+ {\cal O}(1/t) \right)
 \nonumber\\
&& \quad \times \frac{b_{n}(\{x_{i}\})}{(2 \pi)^{n/2}n !}
\int d^{n}y \ e^{-\sum y_{i}^{2}/2t}
\prod_{1 \leq i < j \leq n} |y_{j}-y_{i}|
\nonumber\\
&& = t^{-n(n-1)/4} \left( 1+ {\cal O}(1/t) \right) 
 \nonumber\\
&& \quad \times \frac{b_{n}(\{x_{i}\})}{(2 \pi)^{n/2}n !}
\int d^{n}u \ e^{-\sum u_{i}^{2}/2}
\prod_{1 \leq i < j \leq n}
|u_{j}-u_{i}|,
\nonumber
\end{eqnarray}
where $u_{i}=y_{i}/ \sqrt{t}$.
The last integral is the special case 
($\gamma=1/2$ and $a=1/2$) of 
\begin{eqnarray}
&&\int d^{n} u \ e^{- a \sum u_{i}^{2}}
\prod_{1 \leq i < j \leq n}
|u_{j}-u_{i}|^{2 \gamma} 
\nonumber\\
&& \qquad = (2 \pi)^{n/2}
(2a)^{-n(\gamma(n-1)+1)/2} 
\prod_{i=1}^{n} \frac{\Gamma(1+i \gamma)}{\Gamma(1+\gamma)},
\nonumber
\end{eqnarray}
which was derived by Mehta (eq.(17.6.7) on page 354 
in \cite{Meh91}) as a consequence of Selberg's integral.
Here $\Gamma(x)$ is the Gamma function with the values
$\Gamma(3/2)=\sqrt{\pi}/2$ and
$
\prod_{i=1}^{n} \Gamma(1+i/2) = 2^{-n(n-1)/2} 
(\sqrt{\pi}/2)^{n} n ! \ a_{n},
$
where $a_{n}$ is given by (\ref{eqn:Krat2}).
Then we have 
\begin{eqnarray}
&&{\cal N}_{n}(t; \{x_{i}\}) \nonumber\\
&& =
t^{-n(n-1)/4} 2^{-n(n-1)/2} 
a_{n} b_{n}(\{x_{i}\}) \left( 1+ {\cal O}(1/t) \right)
\qquad 
\label{eqn:Brown}
\end{eqnarray}
as $t$ tends to infinity.
Using (\ref{eqn:Brown}) 
the asymptotic form in $ t \to \infty$ 
of the probability density is given as
\begin{eqnarray}
&&g_{n}(t; \{y_{i}\}| \{x_{i}\})=
{\bf 1}\{y_{1} < y_{2} < \cdots < y_{n} \} 
c_{n} t^{-n(n+1)/4}
\nonumber\\
&& \quad \times e^{- \sum y_{i}^{2}/2t} 
\prod_{1 \leq i < j \leq n}
(y_{j}-y_{i}) \
\left(1+ {\cal O}(1/t) \right)
\nonumber
\end{eqnarray}
with
$c_{n}=2^{n(n-1)/2}/\{(2 \pi)^{n/2} a_{n}\}$
and thus
\begin{eqnarray}
&&\lim_{t \to \infty} d \mu_{n}
(t; \{\sqrt{t/\tau} u_{i}\}|\{x_{i}\}) =
{\bf 1}\{u_{1} < u_{2} < \cdots < u_{n} \} 
\nonumber\\
&& \ \times c_{n} \tau^{-n(n+1)/4} \ 
e^{ - \sum u_{i}^{2}/ 2 \tau }
\prod_{1 \leq i < j \leq n}
(u_{j}-u_{i}) 
d^{n}u \qquad
\label{eqn:limit1}
\end{eqnarray}
for any $\tau > 0$.
We note that the diffusion scaling limit,
$L \to \infty$ with $m=Lt, s_{i}= \sqrt{L} x_{i}/2,
e_{i}= \sqrt{L} y_{i}/2$,
and the above $t \to \infty$ limit with 
$y_{i}=\sqrt{t/\tau} u_{i}$ are combined to define a limit
$L^{\prime} \equiv Lt/\tau \to \infty$ with 
$m= L^{\prime} \tau, 
e_{i}=\sqrt{L^{\prime}} u_{i}/2$
and $s_{i}$ in the smaller order than $\sqrt{L^{\prime}}$.
Since $s_{i}/e_{i} \to 0$ in this limit,
the limit (\ref{eqn:limit1})
should be regarded as the scaling limit
$d \mu_{n}(t; \{y_{i}\}|\{x_{i}\})$ for
$x_{i}=0, i \in I_{n}$.
That is, when all walkers are started from 
the origin, the probability density
in the scaling limit should be
\begin{eqnarray}
&&g_{n}(t; \{y_{i}\}|\{0\})
= {\bf 1}\{y_{1} < y_{2} < \cdots < y_{n} \}
\ c_{n} t^{-n(n+1)/4}
\nonumber\\
&& \qquad \times 
\exp \left(- \frac{1}{2t} \sum_{i=1}^{n} y_{i}^{2} \right)
\prod_{1 \leq i < j \leq n}
(y_{j}-y_{i}).
\label{eqn:limit2}
\end{eqnarray}
By using the scaling property (\ref{eqn:scaling}),
we can give another explanation for the reason 
why the above limit procedure gives 
the case for $x_{i} \equiv 0$ as follows.
The scaling property (\ref{eqn:scaling}) can be written 
for distribution functions as
$$
d \mu_{n}(t; \{y_{i}\}|\{x_{i}\})
= d \mu_{n}(\ell^2 t; \{\ell y_{i}\}|\{\ell x_{i}\})
$$
for any integer $\ell$. Then, if we set
$x_{i} \equiv 0$ and $T= \ell^2 t$,
we have the invariance
\begin{equation}
d\mu_{n}(t; \{y_{i} \}|\{0\})
= d\mu_{n}(T; \{\sqrt{T/t} \ y_{i}\}|
\{0\})
\label{eqn:inv}
\end{equation}
for any $T > 0$. Then we can take the limit 
$T \to \infty$, which will derive the expression 
(\ref{eqn:limit2}).

The result (\ref{eqn:limit2}) is very important, since
if we set $t=1$
and assume that
$y_{1} < \cdots < y_{n}$,
then we have the equality
$$
g_{n}(1; \{y_{i}\}|\{0\})
= n ! \ g_{n}^{{\rm GOE}}(\{y_{i}\})
$$
where $g_{n}^{{\rm GOE}}(\{y_{i}\})$ is the
probability density of eigenvalues
of random matrices in the GOE \cite{Meh91}.
The equality (\ref{eqn:inv}) with $T=1$ implies that
$d \mu_{n}(t; \{y_{i}\}|\{0\})$
is a function of $n$ variables $\{u_{i}=y_{i}/\sqrt{t}\}$. 
In other words, the dynamical scaling with
the classical exponent $\theta=2$ 
(see, for example, chapter 16 in \cite{Hen99})
is established, in which the scaling function is 
exactly and explicitly given by 
$n ! \times g_{n}^{{\rm GOE}}(\{ u_{i}\})$
with the scaling variables $u_{i}=y_{i}/t^{1/\theta}$.

In the present diffusion scaling limit, 
the random walks converge to the Brownian motions
in distribution.
Then (\ref{eqn:Brown}) gives the asymptote in $t \to \infty$ 
of the probability that the $n$ independent one-dimensional
Brownian motions started from 
$x_{1} < x_{2} < \cdots < x_{n}
\ (x_{i} \in {\bf Z})$ do not intersect
up to time $t$.
The scaling property (\ref{eqn:scaling}) is nothing but the
diffusion scaling of Brownian motions,
$
{\cal N}_{n}(\ell^2 t; \{\ell x_{i}\})
= {\cal N}_{n}(t; \{x_{i}\}).
$
Moreover, we can define a diffusion process 
by giving the transition probability density as
\begin{eqnarray}
&&p_{n}(s; \{y_{i}\}|\{x_{i}\}) =
\lim_{t \to \infty} \frac{f_{n}(s; \{y_{i}\}|\{x_{i}\})
{\cal N}_{n}(t; \{y_{i}\})}
{{\cal N}_{n}(t+s; \{x_{i}\})} \nonumber\\
&& \quad =
\frac{b_{n}(\{y_{i}\})}{b_{n}(\{x_{i}\})}
f_{n}(s; \{y_{i}\}|\{x_{i}\}),
\nonumber
\end{eqnarray}
where (\ref{eqn:Brown}) has been used \cite{KT}.
Isozaki and Yoshida \cite{IY} studied the $n=2$ case
as the limit process of the two friendly walkers in 
the ``wetting phase".

In summary, we have performed the diffusion scaling limit 
of the vicious walkers and determined the time-dependent
spatial-distribution functions in the limit.
There the Schur function plays
an important role to describe the initial-configuration
dependence.
The relation with the
Gaussian orthogonal ensemble of random matrices
was clarified.
In the present study we have constructed 
new continuous-time processes on the real axis 
for arbitrary numbers of walkers $n$. 
We remark that the results reported in this Letter
may hold also in the fully ``wetting phases" \cite{Brazil} 
of the models of friendly walkers \cite{TK98,CC99,GV,IK01}, 
which are related with a famous unsolved problem, 
the directed percolation problem.
Further study of the distribution functions and
diffusion processes derived in this Letter will be desired.

The authors acknowledge useful discussions with
J. Cardy, D. Arrowsmith and J. Essam.


\begin{thebibliography}{99}
\bibitem{Schur}
A partition 
$\lambda=(\lambda_{1}, \lambda_{2}, \cdots, \lambda_{n})$
is a non-increasing series of non-negative integers,
$\lambda_{1} \geq \lambda_{2} \geq \cdots \geq \lambda_{n}
\geq 0$.
The Schur function 
$s_{\lambda}(z_{1}, \cdots, z_{n})$
specified by $\lambda$ is
a polynomial of $(z_{1}, \cdots, z_{n}) \in V
\equiv$ the $n$ dimensional complex vector space,
defined by 
$ \displaystyle{ 
s_{\lambda}(z_{1}, \cdots, z_{n}) =
\det_{1 \leq i, j \leq n} 
\left(z_{i}^{\lambda_{j}+n-j}\right)/
\det_{1 \leq i, j \leq n} \left(z_{i}^{n-j}\right).
}$
It is a symmetric polynomial of degree 
$\sum_{i=1}^{n} \lambda_{i}$ in $z_{1}, \cdots, z_{n}$ and
it is know that (see p.44 in \cite{Mac95})
$
s_{\lambda}(1,1, \cdots, 1)=
\prod_{1 \leq i < j \leq n}
(\lambda_{i}-\lambda_{j}+j-i)/(j-i).
$
The Schur function is a character of the irreducible 
representation specified by $\lambda$ of the group $GL(V)$
and its value at $z_{1}=\cdots=z_{n}=1$ is the dimension
of the representation 
(see, for example, Lecture 6 in \cite{FH91}).
\end{thebibliography}

\end{document}